\documentclass{sigcomm-alternate}
\usepackage{graphicx}
\usepackage{url}
\usepackage{booktabs}
\usepackage{mathptmx}
\usepackage{color}
\usepackage{caption}
\usepackage{subcaption}

\begin{document}

\title{Approximate Networking for \\Global Access to the Internet for All (GAIA)} 
\author{\large Junaid Qadir$^{1}$, Arjuna Sathiaseelan$^{2}$, Liang Wang$^{2}$, Barath Raghavan$^{3}$\\
\normalsize $^{1}$Information Technology University (ITU)-Punjab, Lahore, Pakistan\\
\normalsize $^{2}$Computer Laboratory, University of Cambridge, United Kingdom\\
\normalsize $^{3}$ICSI, Berkeley, USA. \\
\normalsize Email: junaid.qadir@itu.edu.pk; arjuna.sathiaseelan@cl.cam.ac.uk; liang.wang@cl.cam.ac.uk; barath@icsi.berkeley.edu}
\maketitle

\begin{abstract}

Decades of experience have shown that there is no single one-size-fits-all solution that can be used to provision Internet globally and that invariably there are tradeoffs in the design of Internet. Despite the best efforts of networking researchers and practitioners, an ideal Internet experience is inaccessible to an overwhelming majority of people the world over, mainly due to the lack of cost efficient ways of provisioning high-performance global Internet. In this paper, we argue that instead of an exclusive focus on a utopian goal of universally accessible ``ideal networking'' (in which we have high throughput and quality of service as well as low latency and congestion), we should consider providing ``approximate networking'' through the adoption of context-appropriate tradeoffs. Approximate networking can be used to implement a pragmatic tiered global access to the Internet for all (GAIA) system in which different users the world over have different context-appropriate (but still contextually functional) Internet experience. 


\end{abstract}

\section{Introduction}

Internet access is a key indicator of the potential of economic progress. Internet's impact is imprinted on all spheres of human life---personal, societal, political, economical,  and educational---in both developing and developed countries. The fact that Internet access can play a large role in facilitating development motivates the vision of \textit{Global Access to the Internet for All} (GAIA), currently being formally pursued in the Internet Research Task Force (IRTF). Bringing the Internet to the remaining billions of people will democratize knowledge, open up new opportunities, and undoubtedly open up avenues for sustained development. 

While Internet has the capability of fostering development and growth, this potential is thwarted by the inability of billions of people to access the Internet. According to recent statistics, almost 6 billion people do not have high-speed internet, which makes them  unable to fully participate in the digital economy \cite{worldbank2016}. To be sure, this ``digital divide'' is not a binary divide. There is a whole spectrum of connectivity options and digital capabilities that can be accessed by human beings around the world. In some places, ultra high-speed broadband connections are available, while there are hosts of places where there is no connectivity at all---the majority, however, lie somewhere in between. 

To deliver universal digital access, no single solution appears to fit all needs. Despite its great success, cellular technology cannot alone enable GAIA (since it is mainly an urban phenomena that cannot be used to cost effectively serve rural and remote areas \cite{subramanian2006rethinking}). Similarly, other technologies---such as millimeter wave wireless (mmWave), satellite services, high-altitude platforms such as Google Balloon, solar planes/ drones such as the Aquila drone proposed by Facebook---have their own pros and cons. A more promising solution will be to develop an all-inclusive architecture that can choose the right tradeoff point according to the situation.  

\subsection{What is Approximate Networking?}


While ideally speaking, we will like an Internet that is perfect, and has extremely high capacity, bandwidth, and reliability in addition to extremely low or negligible delays, errors, and congestion. For practical purposes, the modern fiber-based broadband high-speed networks available in select places (mostly in advanced countries) come close to this ideal. We call such networks ``ideal networks''. In contrast, we consider ``\textit{approximate networks}'' that are networks that make some design tradeoffs to deal with different challenges and impairments. We note here that ideal networks and approximate networks do not define a binary divide but a spectrum of options. Approximate networking is the idea that protocols and architectures of network systems can let applications trade off service quality for efficiency in terms of cost/ affordability/ accessibility. We can also define approximate networks as networks that come close to ideal networks in quality, nature, and quantity.\footnote{Oxford Dictionary: Approximate (v): come close or be similar to something in quality, nature, or quantity.}

\subsection{Why Adopt Approximate Networking?}
\label{sec:motivations}

\vspace{.5mm}
``\textit{Be approximately right rather than exactly wrong}.''---Tukey.
\vspace{.5mm}

We need ``approximate networking'' when the imperfections of the real world preclude an ``ideal networking'' solution. We argue that approximate networking is not only useful but also inevitable. Approximate networking is appropriate when any of the ideal networking  assumptions---e.g., that there is 24x7 connectivity; an end-to-end path is always available; the latency is never too high (i.e., is less than (half) a second); the network does not have a high error rate---are not met. In a world where even the specifics of an ideal network are dynamic (even as newer higher-capacity technologies emerge, bandwidth-hungry applications are sprouting up even more rapidly\footnote{Cisco Visual Networking Index, White Paper, Feb 2015}), achieving the goal of universal ``ideal networking'' appears a Sisyphean task. Some important reasons we should seriously consider approximate networking are described next.

\vspace{.5mm}
\subsubsection{Affordable Universal Internet (GAIA)}


The right of affordable access to broadband Internet is enshrined in the 2015 sustainable development goals of the United Nations. The ITU broadband ``Goal 20-20'' initiative aims at an optimistic target of universal broadband Internet speeds of 20 Mbps for \$20 a month, accessible to everyone in the world by 2020 (Source: Alliance for Affordable Internet (A4AI) Report, 2014). Such an approach---that aims at providing an ``ideal networking'' experience universally---has historically always failed (due to various socioeconomical and technical issues). An important reason is that most modern technologies (such as 3G/ 4G LTE and the planned 5G) are urban focused since rural systems (being sparsely populated by definition) do not thus hold much business potential for mobile carriers \cite{subramanian2006rethinking}. The current percentage of households with Internet access worldwide is 46.4\%, with only 34\% and 7\% of the households in developing countries and least-developed countries having Internet access, respectively \cite{WDR2016}. The Internet is also large unaffordable when we consider that on average the mobile broadband price and the fixed-line broadband prices are 12 and 40\% of the average person's monthly income (with women and rural populations hit the most).\footnote{Source: A4AI survey conducted in 51 countries in 2014.} The UN target aims to reduce the true cost of connecting to the Internet to around 5\% of a person's monthly income. Approximate networking is a particularly appealing option to reach out to the offline human population (more than 4 billion of which live in developing countries).

\subsubsection{The Pareto Principle (80-20 Law): The Power of ``Good Enough''}

To help manage the approximate networking tradeoffs, it is instructive to remember the \textit{Pareto principle}, alternatively called the \textit{80-20} rule \cite{koch201180}---which states that roughly speaking that 20\% of the factors result in 80\% of the overall effect. This principle has big implications for approximate networking since this allows us to provide adequate fidelity to ideal networking by only focusing on the most important 20\% of the effects. Alternatively put, this theory states that 80\% of what goes into creating the ideal networking experience provides little cosmetic benefits to the user. The key challenge in approximate networking then becomes the task of separating the all-important essential non-trivial factors from the trivial factors (which may be omitted or approximated). In this regard, we can leverage previous human-computer-interaction (HCI) research that has shown that human quality of service (QoS) perception can be flawed (e.g., relatively fast service may be judged to be unacceptable if the service is not predictable, visually appealing and reliable \cite{bouch2000quality}) in choosing the the precise approximate networking tradeoff to adopt so that the users perceive the least inconvenience. 

\subsubsection{Need of Energy Efficiency}

Information and communication technology (ICT) is a big consumer of world's electrical energy, using up to 5\% of the overall energy (2012 statistics) \cite{raghavan2011networking}. The urgency of delivering on the front of energy efficiency is reinforced by the impending decline of non-renewable energy resources along with the concomitant increase in ICT demand. The approximate networking trend can augment the hardware-focused ``approximate computing'' trend in managing the brewing energy crisis through the ingenuous use of approximation. Through its relevance for energy efficiency, approximate networking is suitable for both ICTD research (that focuses on developing countries) and for research in the LIMITS context \cite{raghavan2011networking} that focuses on ``undeveloping countries''\footnote{The moniker of ``\textit{undeveloping countries}'' refers not only to those developing countries that face degradation of their social and economic systems but also subsumes developing countries} \cite{approximateLIMITS}. 
 
\subsection{Approximation in Related Domains}
   
Approximate networking is the networking analog of the emerging computer architecture trend called ``\textit{approximate computing}'' \cite{han2013approximate} in which approximations are performed at the hardware level to boost the energy efficiency of systems. In the world where Moore's law breaks down (due to physical limits such as heat), new approaches will be inevitably needed to the growing demand to process ever-increasing amounts of data \cite{nair2015big}. Broadly speaking, approximate computing leverages the capability of many computing systems and applications to tolerate some loss of quality and optimality by trading off ``precision'' for ``efficiency'' (however, these may be defined). In computer science, there are many applications that are imprecision tolerant (in particular, this tolerance is a familiar hallmark of image and video processing). It has been shown that by relaxing the need of exact operations, approximate computing techniques can significantly improve energy efficiency \cite{han2013approximate}.

The general idea of approximation is widely used in various diverse sciences and technologies. We use approximation in measurement (all measurement devices have finite precision) and in digital computing (since real numbers cannot be stored on finite precision computing devices). We use approximation in multimedia communication where slightly inaccurate data can barely be perceived by the users but can significantly improve the network's efficiency. We use approximation when the problem does not admit an efficient optimal solution: in such cases, we lower down our targets from optimizing to \textit{satisficing} (i.e., producing sufficiently satisfactory answers). Despite their approximate nature, these approximations serve us well for most practical purposes. 

\subsection{Approximate Networking: Old Wine in a New Bottle?} 

 Approximation is a classic tool that is employed in computing and networking when faced with constraints, intractability, and tradeoffs. We do not claim that approximate networking is a novel way of dealing with networking problems that have resource constraints. Indeed, delay-tolerant networking (DTN), information-centric networking (ICN), approximate computing, the use of caching and opportunistic communication are all approximate networking solutions. However, we contend that the idea of viewing approximate networking as first-class networking and to view it as an overarching general framework for dealing with context-appropriate networking tradeoffs holds new promise. The concept of approximate networking generalizes the concepts of lowest-common denomination networking (LCD-NET) \cite{sathiaseelan2013lcd} in that the idea of approximate networking extends to the design of network infrastructure as well as algorithms and protocols. Another related research theme is that of \textit{challenged networks}---which are networks that have very long communication delay or latency; or unstable or intermittently available links; or very low data rates; or very high congestion; or very high error rate.

\subsection{Approximate Networking: Is it Common?} 

\vspace{2mm}
``\textit{All models are approximations. Essentially, all models are wrong, but some are useful.}''---George Box.
\vspace{2mm}


Taking a broad view, we see that many established existing technologies are in fact examples of approximate networking. The TCP transport layer protocol implements a very useful approximation: it can approximate a reliable network even when none exists. The UDP protocol approximates the transport service provided by TCP but it tradeoffs reliability for performance gains. Virtual networks are approximate networks---but they can, for all practical purposes, serve the needs of users/ applications as well as physical networks can. We can also have approximate networks that provide only a tenuous approximation of the quality, nature, or quantity of the Internet: e.g., services that rely on data mules (e.g., DakNet \cite{pentland2004daknet}) are only infrequency connected to the Internet; there other also services (such as Outernet\footnote{\url{https://en.wikipedia.org/wiki/Outernet}} and Internet in a Box \cite{tyson2015could}) that approximate the Internet experience without actually connecting to the Internet. 



\subsection{Contributions of this paper}


Approximate networking is neither a singular standalone technique nor is it a new way of looking at networking problems that have resource constraints. Indeed, a number of existing networking techniques utilize approximation and best effort. The main contribution of this paper is to propose approximate networking as an overarching framework for systematically thinking about managing networking tradeoffs. Approximate networking can also guide us about which tradeoff to adopt in a given situation depending on the context and our goals. We show that approximate networking is not a niche topic that relates only to ICTD research, but that it has has general implications for networking (not only for the design of network architectures but also for network protocols and algorithms). In this paper, we emphasize that for universal Internet provisioning of mobile and Internet services, it is time to move away from pursuing overengineered ``perfect products'' and focus instead on developing appropriate ``good enough'' solutions.  We also highlight that ``good enough'' solutions do not simply mean stripped-down versions of existing high-end products (which is a short-term solution that will only leave customers feeling shortchanged) but will require the creation of affordable solutions from the ground up to deliver higher value to customers.

\section{Approximate Networking\\Technologies}

\subsection{Approximate Networking Hardware}

Digital computer are necessarily approximate (since real numbers cannot be represented in digital computers). With the rise of big data, and the increase in the throughput and computing prowess of modern networking devices, the energy efficiency has become a major concern, especially when we consider the warehouse scale of modern datacenters  \cite{nair2015big}. Future generation networking systems will have to look towards approximate computing techniques since modern computers are now reaching physical limits regarding the energy efficiency of miniaturized electronics. 

The general trend of trading off some accuracy for increase in performance is known as approximate computing. The approximate computing solution has been proposed as a ``good enough'' solution \cite{kugler2015good} that can deal with this energy crunch by trading off the accuracy of computation for gains in energy and performance. There is a rich broader tradition amongst hardware designers to design algorithms optimized for exploiting the hardware wherever possible (e.g., by designing algorithms that perform computations in powers of 2, even when that leads to minor, but inconsequential, errors; see more examples in \cite{varghese2010network}). 


\subsection{Approximate Networking Software}


The idea of approximation is an oft-used tool in networking \textit{algorithms} and \textit{protocols} \cite{varghese2010network}. Approximate networking \textit{algorithms} (also called heuristics) are often required in networking to tackle discrete optimization problems (many of which are NP-hard, and thus there are no efficient algorithms to find optimal solutions). Such algorithms have been widely used in scheduling, routing, QoS problems in networking. 

In particular, Bloom filters---which is a high-speed approximate set membership query algorithm tests that can return false positives (but never false negatives)---have been extensively applied in networking in a wide variety of settings \cite{broder2004network} (such as caching, peer-to-peer systems, routing and forward, and monitoring and measurement). Bloom filters are important since a large number of applications require fast matching of arbitrary identifiers to values, and with millions (or even billions) of entries being common, scalable methods for storing, updating, and querying tables are required. Broder and Mitzenmacher have articulated the \textit{Bloom filter principle} \cite{broder2004network}: ``Whenever a list or set is used, and space is at a premium, consider using a Bloom filter if the effect of false positives can be mitigated''. 


A number of existing \textit{protocols} can be envisioned as approximate networking solutions (e.g., a protocol may adopt an ``approximation'' by (1) delivering frames that have errors to the upper layers; (2) delivering duplicate packets; (3) delivering packets that may have violated some security policy). These seemingly suboptimal decisions can be adopted for some other convenience (such as affordability, cost, performance, efficiency). 

The most common type of approximate networking involves delivering data to the upper layer that may have uncorrected errors. For example, reliable transport protocols like TCP are too unwieldy for multimedia applications (which require speedy transfer, even if it has some error). UDP can be considered as an approximate networking solution since it trades off reliability for efficiency and rapidness. UDP-Lite \cite{larzon2004lightweight} takes the approximation further by adopting \textit{partial checksums} (rather than full checksums as in UDP/TCP) to cater to multimedia applications whose performance degrades when erroneous packets are discarded because of bit errors (as usually happens in TCP/ UDP, which discards packets that fail the checksum). 

Approximation networking can also be used to provide ``good enough'' networking services in intermittently-connected networks. In intermittent networks, nodes route data opportunistically towards the destination when appropriate relays have an encounter. The de facto routing protocol for the DTN community is the \textit{Bundle protocol} \cite{scott2007bundle} that has a number of implementations. The Bundle protocol can take advantage of scheduled, predicted, and opportunistic connectivity in addition to continuous connectivity and makes use of loose coupling. DTN-specific application-layer protocols have also been developed that can provide file transfer, streaming, and multicast service \cite{burleigh2003delay}. 
 
\section{Principles for Approximate Networking}

\vspace{.5mm}
``\textit{The principle of constant change is perhaps the only principle of the Internet that should survive indefinitely}''---RFC 1958.
\vspace{.5mm} 

Internet design principles have always been influential in the design of its protocols \cite{clark1988design}. With the Internet evolving in its architecture, users, and applications, the design principles of the Internet also need to evolve. While it is daunting to state the principles of a dynamic complex artifact such as the future Internet, we will describe in this section the most important five principles that can act as steering principles for approximate networking. The proposed principles draw from other principles proposed in literature (such as principles proposed for mobile networking in \cite{scott2006haggle}, computing within limits \cite{raghavan2012intermittent} \cite{chen2015computing}, and for frugal innovation under scarcity and austerity  \cite{radjou2012jugaad} \cite{mullainathan2013scarcity}). We explain these principles next.






\subsection{P1---Being Dynamic \& Flexible}

\vspace{.5mm}
``\textit{Be like water}''---Lao Tzu.
\vspace{.5mm}


In approximate networking, instead of relying on predetermined paths and methodologies, it makes better sense to adopt dynamic and flexible communication where possible in which resources are allocated dynamically and any unused capacity is reclaimed or ``scavenged''. There are many trends that follow this principle including packet switching, statistical multiplexing, dynamic spectrum access and less-than-best effort (LBE) service through scavenger access \cite{shalunov2001qbone}. Adopting dynamic resource allocation can help mitigate inflexible static resource allocation by allowing some extra slack in dealing with scarcity \cite{mullainathan2013scarcity}. Being flexible encompasses the use of non-conventional data transfer methods \cite{scott2006haggle}, such as \textit{neighborhood connectivity}: e.g., as used in community networking or when devices connect to other local devices; and \textit{user or device mobility} that can physically carry data: e.g., user mobility that can be used to physically transfer significant amounts of data from one place to another (using approaches such as \textit{data mules} or \textit{message ferrying} \cite{scott2006haggle}). In addition, approximate networks can benefit from the flexibility of temporal shifting of elastic traffic to allow load balancing and mitigate peak time congestion and utilize the off-peak spare capacity.

\subsection{P2---Keeping the Architecture Decoupled}


\vspace{.5mm}
``\textit{Modularity is good. If you can keep things separate do so}''---RFC 1958.
\vspace{.5mm}

The remarkable success of the Internet protocols that have survived for more than 40 years may be attributed to the forward-looking philosophy of the original Internet designers who ``emphasized loose coupling'', which has allowed the Internet to: use a rich variety of underlying media; support independent policies (according to autonomous systems); and be more resilient \cite{cerf2013loose}. Where the Internet design missed this trick---notably in the coupling of the host identifier and location in IP addressing---the result is inflexibility in coping with new applications (e.g., user mobility). In recent times, researchers have proposed many architectures---such as delay-tolerant networking (DTN) and information-centric networking (ICN)---that aim to redress such couplings and thereby provide a GAIA framework \cite{trossen2016towards}. It has long been known that the marriage of wireless and asynchronous service can be kernel around which a universal broadband solution may be built \cite{pentland2004daknet} and that \textit{connectionless/ asynchronous communication model} is sufficient for most rural community needs. The Publish/ Subscribe (Pub/Sub) model implemented in many ICN/DTN architectures naturally allows asynchronous communication and \textit{space/time decoupling}, which is especially relevant in environments where end-to-end communication may not be possible.  

\subsection{P3---Resource Pooling \& DIY Networking} 

\vspace{.5mm}
``\textit{Innovative bottom-up methods will solve problems that now seem intractable---from energy to poverty to disease}.''---Vinod Khosla
\vspace{.5mm}

Broadly speaking, resource pooling involves abstracting a collection of networked resources to behave like a single unified resource pool, and developing mechanisms for shifting load between the various parts of the resource pool. The main benefits of resource pooling include greater reliability and increased robustness against failure; better ability to handle surges in load on individual resources; and, increased utilization \cite{wischik2008resource}. Resource pooling is well suited for the ``undeveloping world'', where maintaining dedicated infrastructure is especially cost prohibitive for small-scale entrepreneurs, business owners, and non-profits. Resource pooling can be especially influential in scarcity-afflicted approximate networking settings since resource pooling naturally allows some slack in dealing with with scarcity and failures \cite{qadir2016resource}. The general idea of resource pooling encompasses modern trends of dynamic spectrum access networks (that utilize the so called spectrum ``white space''), community networks, multihoming with heterogeneous technologies, network coding, and multipathing. 

Amongst the various resource pooling techniques, community networking---being a do-it-yourself (DIY) cost-effective networking solution---is especially promising for approximate networking. To facilitate crowdsourcing, a number of open-source projects (e.g., Haggle \cite{scott2006haggle}) have emerged that can be leveraged for building autonomous neighborhood networking. In recent times, it has even become possible to develop \textit{community cellular} networks using low-cost software defined radios (SDRs) and open-source software such as OpenBTS \cite{heimerl2013local}. Such community-driven projects can be used to provide approximate networking services where traditional ideal networking solutions are not feasible (e.g., in rural settings). 

It has been shown in literature that keeping a margin or keeping some slack is a key to frugal innovation \cite{radjou2012jugaad} and thriving in scarcity-afflicted environments \cite{mullainathan2013scarcity}. In particular, approximate networking solutions can leverage the inherent diversity and multiplicity of networks to reap the benefits of increased reliability, efficiency, and fault tolerance \cite{qadir2015exploiting}. The ideas of multiplicity, redundancy, and using slack---which may appears out of place for approximate networking---is, counterintuitively, extremely important since any approximate networking solution for challenging environments that does not have redundancy and slack inbuilt will be debilitatingly fragile. 

\subsection{P4---Failure Cognizant Network Design} 

\vspace{.5mm}
``\textit{Hoping for the best, prepared for the worst, and unsurprised by anything in between.}''---Maya Angelou.
\vspace{.5mm}

%

Approximate networking should be designed by assuming an inevitable presence of failures/ weaknesses/ deficiencies. Failures should be anticipated, and even intentionally utilized where appropriate (e.g., the ``random early detection'' (RED) congestion control algorithm intentionally drops some packets when the average queue buffer lengths are more than a threshold (e.g. 50\%) to implicitly signal to the sender about the rising congestion). Approximate networks typically have to deal with high bit error rates (BER); very low/ variable bandwidth; long signal propagation delays; unstable or intermittently available links; and high congestion. The design space should explicitly consider what tradeoffs should be adopted according to the context of the user and application (Section \ref{sec:tradeoffs}). 



The principle about designing for failure should not be construed to mean that approximate networking should not try to avoid failure. To the contrary, it is very important for approximate networking solutions to be robust\footnote{We define some property of a system to be robust if it is invariant with respect to some set of perturbations. Fragility is defined as the opposite of robustness.}  and to fail gracefully when subsystems fail. Approximate networking solutions should aim to avoid disruption due to failures by adopting robust tradeoffs that make the solution failure proof or resilient. Towards this end, it has been pointed out in literature that the solutions should ``keep the margin'' \cite{radjou2012jugaad} and should have ``spare bandwidth'' \cite{mullainathan2013scarcity} when working in scarcity-afflicted environments. 

\subsection{P5---Scarcity Inspired Network Design}

\vspace{.5mm}
``\textit{One cannot alter a condition with the same mindset that created it in the first place.}''---Albert Einstein
\vspace{.5mm}

Scarcity inspired network design aims at \textit{doing more with less}. In particular, such design aims to built scarcity aware networking solutions and avoid \textit{indifference} (in which a fully-featured ideal networking product is used unmodified in a challenging environment) and \textit{defeaturing} (in which a fully-featured ideal networking product is stripped of some nice-to-have features and then implemented in approximate networking settings). Both these practices (i.e., indifference and defeaturing) are doomed to fail in approximate networking setting since the product's underlying design does not account for the inherent fundamental constraints or the socioeconomic context of the environment. 


Apart from the fact that the majority of the people in developing world do not have Internet access, people who do get online are often encumbered by poor network connectivity, and prohibitively slow/ unstable services. There are a number of reasons why the conventional protocols and services are ill-suited for such challenging environments \cite{chen2015computing}. This motivates the development of optimized protocols and services that can work well in such poor-connectivity scenarios. An effective approach in such settings is to adopt a design that emphasizes simplicity. Simplicity has always been considered a virtuous design trait---e.g., this has been codified in the engineering principles of KISS (``Keep it Simple, Stupid'') and the ``Occam's Razor'' (which recommends adopting the simplest design solution for protocols and not to multiply complexity beyond what is necessary. The design of simple convenient and accessible approximate networking solutions can accelerate the adoption of approximate networking since it has been shown time and again that users are willing to trade off fidelity of user experience to gain on accessibility and convenience. Previous work has shown that simple protocols with severe constrains can still enable ``rich'' applications. For example, in situations where mobile users cannot access data services (e.g., due to services not being offered in that location or due to unaffordability): the users can access services through short messaging service (SMS) and asynchronous voice services \cite{heimerl2009message}.

\section{Tradeoffs Involved in Approximate Networking?}
\label{sec:tradeoffs}

A \textit{tradeoff} refers to the fact that a design choice can lead to conflicting results in different quality metrics. The performance of computers networks depends routinely on multiple parameters and the adoption of tradeoffs is routine. To make matters more complex, the relationship between multiple quality metrics (e.g., throughput and delay) is non-linear \cite{bertsekas1992data}). Since these multiple objectives often conflict with each other, it is rare to find one-size-fits-all solution and tradeoffs have to be necessarily employed. We can borrow concepts from economics to study scarcity and choice. The concept of \textit{opportunity cost}---which is the ``cost'' incurred by going with the current choice and not adopting any other choice---is a key idea that can be used to ensure efficient usage of scarce resources. Another important concept is that of \textit{Pareto optimality}, which refers to a state of resource allocation in which it is not possible to to make any one individual better off without making at least one individual worse off. We can make a Pareto improvement, if we can make at least one individual better off without making any other individual worse off. 

\subsection{What are the main tradeoffs in networking?}



\subsubsection{Latency vs. Throughput}

If high latency can be tolerated, we can achieve extremely high throughput by not using a network at all; but by instead plying trucks chock-full of micro-SD cards containing the data.\footnote{\url{https://what-if.xkcd.com/31/}} This \textit{Sneakernet} concept, long known in networking folklore\footnote{``\textit{Never underestimate the bandwidth of a station wagon full of tapes hurtling down the highway}.''---Andrew Tanenbaum, 1981.}, is the embodiment of the latency-throughput tradeoff. In a similar vein, DTN routing protocols also tradeoff latency for throughput and connectivity---DTN Bundles can achieve the same throughput as IP protocols but with longer latency. It has been shown in literature that throughput-optimal solutions can compromise on delay \cite{bertsekas1992data}. 

\subsubsection{Fidelity vs. Convenience}

A lot of research has shown that customers are willing to sacrifice considerable fidelity for a more convenient and accessible service \cite{maney2010trade}. The notion of fidelity matches with the QoS/ QoE concept but also can include non-tangibles such as social aura and identity.  Convenience refers to the ease of purchase, access, and use; convenience also subsumes concepts such as the cost, accessibility/availability, and simplicity of the service. The disruptive influence of ``good enough technology'', and the user's emphasis on convenience and affordability over perfect quality, can be gauged from an interesting study conducted at the Stanford University that showed that a majority of college students not only were happy with MP3 quality, but actually prefered the average-quality MP3 version of a song played on their iPods to the high-quality CD version \cite{capps2009good}. 


\subsubsection{Performance vs. Cost efficiency}

We can tradeoff \textit{performance (measured in metrics such as resilience, reliability, throughput) to gain on cost efficiency}. 

A cheap way to gain efficiency is to sacrifice resilience and reliability (by employing lesser redundancy) \cite{raghavan2012intermittent}: a scarcely-resourced network will be less costly, but will also have lesser capacity and thus lower throughput for user applications. The tradeoff works the other way as well---we can tradeoff \textit{cost efficiency for increased performance}: i.e., provisioning a better capacity network (by incorporating redundancy) can help improve performance (in terms of throughput and the network reliability/resilience). This is manifested in the ``\textit{multiplexing efficiency vs. QoS}'' tradeoff involved in circuit switching vs. packet switching choice: with circuit switching, voice connections can be served with better QoS---albeit, at a higher cost. 
 

%

\subsubsection{Throughput vs. Coverage/ Reliability}

In wireless networks, there is a tradeoff between the throughput and the coverage (and the reliability) of a transmission---i.e., for higher-rate transmissions, the coverage area is typically smaller, and the chances of BER higher. In certain cases, it may be appropriate to tradeoff coverage for performance, while in other situations, the opposite may be more appropriate.


\subsubsection{Coverage vs. Consumed Power}

In wireless networks, the coverage of a transmission is directly proportional to the transmission power. Since nodes do not need to communicate at all times, researchers have proposed putting to sleep parts of the infrastructure---such as the base transceiver station (BTS) of cellular systems---to save on energy costs. 



\subsubsection{Privacy vs. Free Content/ Services}

While it is usually thought that the Internet services are mostly free, the users trade off their privacy for the right to access Internet services. This tradeoff is becoming more important now with net neutrality and privacy debates becoming more prominent and mainstream. 

\subsubsection{Other Tradeoffs}

There are so many other ways of doing tradeoffs that an exhaustive listing will not be attempted. Many innovative solutions are able to find attractive solutions by inventing a new tradeoff. For example, Vulimiri et al. discovered that an interesting way to reduce latency is to tradeoff some additional capacity or redundancy (i.e., the authors showed that latency can be reduced by by initiating redundant operations across diverse resources and using the first complete response) \cite{vulimiri2013low}. Future approximate networking solutions can derive much utility by focusing on discovering new ways of developing context-appropriate new tradeoffs. 


\subsection{How can we visualize the tradeoffs?}

An interesting approach to understanding tradeoffs is to use visualization techniques. In approximate networking, the task for optimizing for one explicit parameter is easier than optimizing for multiple optimization variables (such as throughput, delay, energy, etc.). The main problem arises when the various objectives---e.g., jointly minimizing both the BER and the transmit power--conflict and and have to compete for dominance. One approach to solving such a problem is to look for a solution on the so called `Pareto Frontier' that defines the set of input parameters that define non-dominated solutions in any dimension. The use of a \textit{tradeoff curve} \cite{van2006trade} can be use to visualize bi-objective problems. The problem of visualizing high-dimensional tradeoffs is more challenging. One approach that has been proposed is to utilize \textit{Pareto front}, which defines the set of values that are each Pareto optimal. 



\begin{figure}
        \centering
        \begin{subfigure}[b]{0.245\textwidth}
                \includegraphics[width=\textwidth]{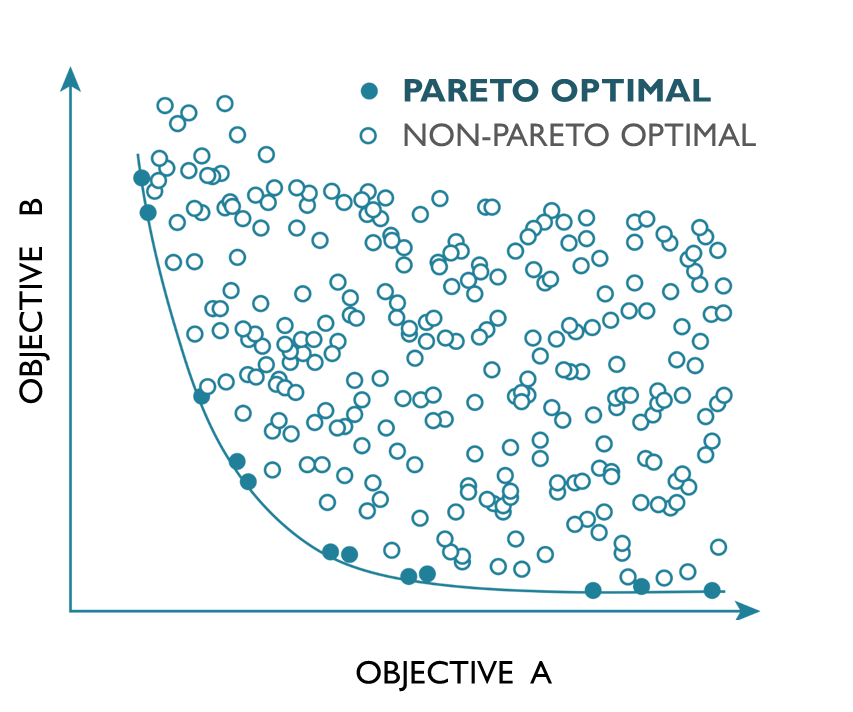}
                \caption{The concept of Pareto frontier}
\label{fig:Paretofrontier}
        \end{subfigure}%
        \begin{subfigure}[b]{0.2\textwidth}
                \includegraphics[width=\textwidth]{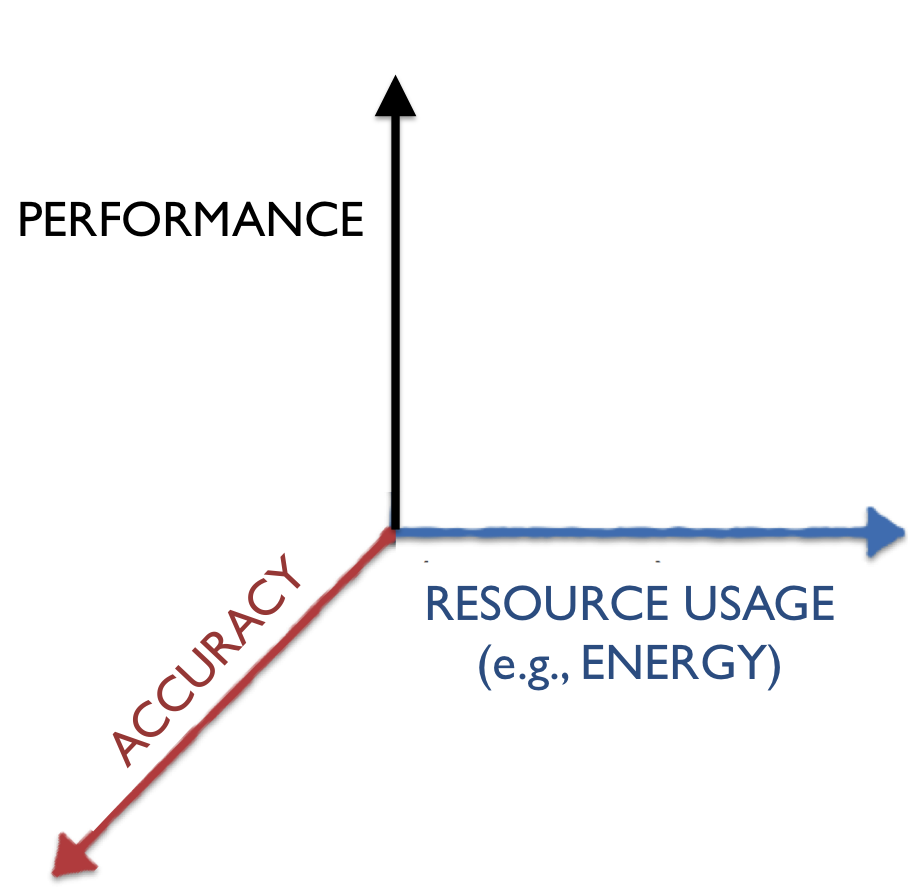}
                \caption{Multi-variate tradeoff}
\label{fig:tradeoff}
        \end{subfigure}
        \caption{How do we define context-appropriate approximate networking tradeoffs?}
\end{figure}

%
%

\subsection{Managing the tradeoffs in networking}

While we have described the main tradeoffs involved in approximate networking and have discussed how they may be visualized, the all-important question still remains to be addressed: \emph{How can we effectively manage these approximate networking tradeoffs?} This is very much an open issue and some open important questions regarding tradeoffs are as follows: 

\begin{enumerate}

\item \textit{How do we quantify when our approximation is working and when it is not?} 

\item \textit{Measuring success in managing the service quality/ accessibility tradeoff?} 

\item \textit{How do we measure the cost of approximation in terms of performance degradation?}

\item \textit{How to dynamically control the approximation tradeoffs according to the network condition.}

\end{enumerate}

The technology-focused concept of \textit{quality of service} (QoS) and the user-focused concept of \textit{quality of experience} (QoE) \cite{bouch2000quality} play an important role in approximate networking tradeoffs. While most QoS works have focused on objective measurable metrics such as delay, jitter, throughput, packet loss etc., both the objective and the subjective quality measures are needed to provide a holistic multi-dimensional assessment. It is important to point out that computing the right tradeoff requires the incorporation of a number of factors such as the subjective user preferences; the subjective and objective user perception of the QoS \cite{bouch2000quality}; the objective application/ service's QoS utility. In addition the tradeoff metrics should also account for the kind of interaction between the various quality metrics (e.g., in a non-zero-sum game, it is not necessary for one metric to lose performance for the other to gain). 

\section{Conclusions}

\vspace{.5mm}
``\textit{What really makes it an invention is that someone decides not to change the solution to a known problem, but to change the question.}''---Dean Kamen
\vspace{.5mm}

The utopian goal of providing ``ideal networking'' service universally is an elusive target (due to the moving target nature of ``ideal networking'' and the lack of affordability of advanced technologies in challenging markets). A lot of experience has highlighted the fidelity-convenience tradeoff according to which users area willing to tradeoff a lot of fidelity for convenience (in terms of accessibility and affordability). In this paper, we have described ``approximate networking'' as a philosophy that understands that there will no one-size-fits-all ideal networking solution that will be universally applicable: approximation networking proposes to adopt appropriate context-specific tradeoffs to provide ``good enough'' service. We have provided an overview of approximate networking technologies and have highlighted how a number of existing Internet technologies can be seen as instances of the larger approximation networking vision. 



\bibliographystyle{ieeetr}

{\scriptsize
\bibliography{approximate}
}


\end{document}